# Customer Identification for Electricity Retailers Based on Monthly Demand Profiles by Activity Sectors and Locations

Joaquin Luque, *Senior Member, IEEE*, Alejandro Carrasco, *Member, IEEE*, Enrique Personal, *Member, IEEE*, Francisco Pérez, and Carlos León, *Senior Member, IEEE*

*Abstract*—The increasing competition in the electric sector is challenging retail companies as they must assign its commercial efforts to attract the most profitable customers. Those are whose energy demand best fit certain target profiles, which usually depend on generation or cost policies. But, even when the demand profile is available, it is in an anonymous way, preventing its association to a particular client. In this paper, we explore a large dataset containing several millions of monthly demand profiles in Spain and use the available information about the associated economic sector and location for an indirect identification of the customers. The distance of the demand profile from the target is used to define a key performance indicator (KPI) which is used as the main driver of the proposed marketing strategy. The combined use of activity and location has been revealed as a powerful tool for indirect identification of customers, as 100,000 customers are uniquely identified, while about 300,000 clients are identifiable in small sets containing 10 or less consumers. To assess the proposed marketing strategy, it has been compared to the random attraction of new clients, showing a reduction of distance from the target of 40% for 10,000 new customers.

## I. Introduction

DURING the last decades, many countries have liberalized in some extent their electricity markets [1]. From a situation where only one or a few state-owned companies were operating, the new regulation process has driven to a scenario with many actors playing different roles along the power delivery chain. Among them, it can be found several production companies that own and manage generation plants. Usually a single transmission system operator (TSO) manages the high-voltage grid. Later, a few distribution system operators (DSOs) are in charge of medium- and low-voltage networks [2]. Finally, the electricity retailers are who carry out marketing, billing, maintenance and customer service activities [3]. An overview of this structure in Europe can be found in [4], [5].

Regarding the retailer level, the competition is especially tough as an increasing number of new companies are being introduced. For instance, Spain has 257 active nationwide retailers (2019 data), becoming the European country with the high number of them and which had experienced the highest increase in this number within the European Union [6].

Electricity retailing can be characterized as a managing risk activity. For medium to long-term (for instance, one year), retailers have to balance two strategic decisions: the procurement of energy in the future trading and the selling price offered to their clients. Also, in the short term, retailers must make decisions associated to the pool trading. As there are uncertainties associated to the prices in the future while the selling prices are fixed, the management of risk is the corner stone of electricity retailing [7]. Then, the actors operating in the electricity retailing arena, focus on optimize their trading portfolio based on a risk assessment, and combining the procurement and selling of energy [8].

Retailers have some alternatives to obtain the electricity, such as spot market, forward contracts, call options or even self-production. They can also finance the investment required for some customers to embrace self-consumption, for instance by installing solar panels, and to tie up that customer for the whole amortization period [9]. With these inputs, retailers can employ several algorithmics methods to decide the optimal selling prices that maximize the benefits in the short and long term [10]. These prices are offered to the customers in several structured tariffs, which consider the cost of producing and distributing the energy and, additionally, send signals to foster more convenient consumptions (for instance, off-peak or greener electricity demand) [11].

The decision-making process of the electricity retailer includes long-term retail load forecasting, power procurement strategies, retail pricing schemes and risk management in the retail market. A survey of these activities and their evolution in recent years can be found in [12]. Several initiatives have also been proposed for an energy retail market strategy that engage companies and consumers in a joint compromise for the transition to a net-zero energy systems, where the electricity is fully produced from renewable and zero-emission sources [13].

Several analytics methods have been proposed to help retailers in their decision-making process, such as electricity customer profiling, demand response and dynamic pricing, or peer-to-peer electricity trading. It is also common to use

This work was supported by the "Ministerio de Ciencia, Innovación y Universidades", Government of Spain under the project "Bigdata Analitycs e Instrumentación Cyberfísica para Soporte de Operaciones de Distribución en la Smart Grid", number RTI2018-094917-B-I00

The authors are with the Department of Electronic Technology, University of Seville, Sevilla, SPAIN (e-mail: jluque@us.es; acarrasco@us.es; epersonal@us.es; fperez@us.es; cleon@us.es)

modelling tools for power system planning, power market simulation, power system simulation, and power consumption analysis [14].

In an increasingly competitive retailing market, the electricity trading companies mainly contend adjusting prices, but also by diversifying products, offering differential prices, expanding channels and deepening cooperation, and using big data analysis to promote users' electricity consumption behavior. The impact of these strategies can be mathematically modelled and assessed [15].

In the retailer's decisions not only technical or economic issues have to be considered. Also, psychological behavior of electricity customers must be anticipated, such as the emotional drivers for selecting a certain retailer company, the reasons for its loyalty to that company, the motivations for leaving, or the willingness to interrupt certain loads under peak demand conditions [16], [17].

Despite the theoretical advantages of a liberalized electricity sector, the retailing market have also shown certain problems. Many newcomers broke, abandoned the market, were taken over, or evolved towards integration into generation for market hedging purposes [18]. The remaining retailers have been forced to dedicate increasing resources to marketing, selling and customer services. This cost is also greater if the current non-negligible churn rate (between 10% to 28% in the Spanish market, depending on the type of customer) [19], is included in the equation. And finally, these costs are partly or fully charged to customers which suffers the contra-intuitive effect that a fierce competition yields higher retailing prices [20].

Then, considering these problems, the economic success of the retailer companies mainly relies in establishing an adequate commercial strategy where benefits can be increased by an optimum selection of the most profitable customers [21]. In this sense, load profiling is the main tool used by retailers to identify target customers [22]–[24]. With this information, different marketing strategies can be set up [15], [16], [25]. These strategies can be based on the hourly [26] and/or the monthly demand profiles [27].

By load profiling it is understood the task of categorizing the electricity demand of a certain customer or set of customers. In the most common scheme, this task is started by a retailer company launching a marketing campaign. Once a potential customer is contacted and identified, its load curve is known besides some personal information (name, address, number of persons in the house, building size, economic activity, etc.). This named information is then compared to the load curves of a large (and anonymous) dataset of customers and it is labelled as belonging to a certain cluster. Then, the corresponding customer profile (assigned cluster) is used for the retailer company to offer individualized marketing proposals to that client, such as pricing schemes, demand-response enrolling propositions, electric vehicle bids, etc. This process, that it is called *named load profiling* in this research, is depicted in the red-shaded left part of Fig. 1. This type of load profiling is performed using several clustering techniques such as K-means, hierarchical clustering, self-organizing map, neural networks, etc. Some reviews of these methods can be found in [28]–[31].

The previously described approach has the drawback of requiring massive and undiscriminated marketing campaigns, either by postal and electronic mails, or by phone calls, which are not only quite costly for the retailer, but also perceived as intrusive and annoying by many potential customers [32]. In this research, this problem is addressed and overcome in a novel way, not found in the literature, for the electricity sector in Spain, by using the information actually available about the economic activity and location linked to every anonymous load profile. In the proposed approach, which is called *unnamed load profiling* and it is depicted in the green-shaded right part of Fig. 1, the load demand of all the customers with the same activity and location is normalized and averaged obtaining a profile for the activity-location pair. Then, the usually very small subset of clients belonging to that pair, is targeted in a more personalized and cost-effective marketing effort.

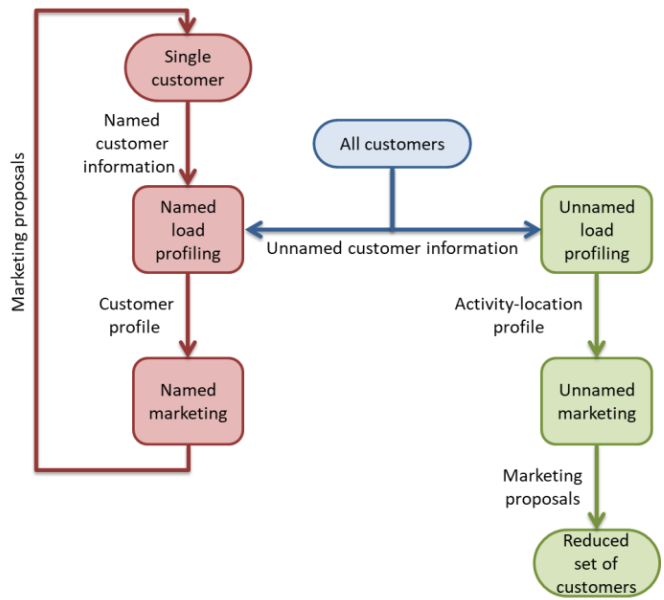

Fig. 1. Load profiling for known (left red-shaded part) and anonymous (right green-shaded part) customers.

Other important novelties of the current work are the large number of customers (more than 27 million) and pairs analyzed (about half a million), the introduction of the enhancement indicator associated to particular customers and to activity-location pairs, and finally, the definition of a marketing strategy for retailing companies.

The remaining of the paper is organized as follows: first, the used dataset is described in Section II.A. Later, up to 3 target profiles are detailed in Section II.B. Then, a metric of the distance between demand and target profiles are introduced in Section II.C. Next, based on this distance, it is defined a key performance indicator (KPI) of the expected enhancement due to selection of a particular customer (Section II.D) or set of customers (Section II.E). Then, the main results obtained applying this methodology are summarized in Section III and discussed in Section IV. Finally, the main conclusions are raised in Section V.

## II. METHODOLOGY

In this section the main definitions and methods to select and assess a retailing marketing strategy will be defined. After describing the datasets employed in the research, different target profiles will be explored. Moreover, the concept of distance between two profiles will be introduced. And finally, an indicator of the advantage of using a certain marketing strategy is proposed.

### A. Datasets

To foster competition in the electricity market, many governments have forced companies to share information about electricity demand of their customers. This information is made available in a semi-public way, that is, it can be accessed for all the registered retailers but not for the general public.

In the Spanish case this dataset contains one record for every customer (up to 27,296,335), anonymously identified by an alphanumeric code, CUPS (Universal Supply Point Code, in Spanish), where each record details the monthly electricity demand of the corresponding customer.

But, by using just that information, it is impossible to reach any particular user and then, no marketing strategy can be addressed based on it. Fortunately, the record corresponding to each customer also contains some additional pieces of information (customer features) originally designed for statistical analysis. Among these features are the following: the contracted power; the location of the user (at municipality level); and its sector of economic activity. The latter is identified by the NACE code (European Classification of Economic Activities, by its acronym in French).

In the case of the electricity market in Spain, 8203 locations are considered, while the economic sector is identified by 1011 different NACE codes. By combining NACE and locations codes up to 8,293,233 different pairs are obtained (1011 × 8203).

Therefore, a new file can be generated including a record for every NACE-location pair. Each record contains a variable-length list of the CUPS that meets the criteria of sharing a same economic activity and location. The individual demand profiles of this set of CUPS can be integrated in an average demand profile of the pair and, as the important information for this work is the shape of the profile, not their absolute values, they are normalized. Finally, some KPIs of every pair are also included, such as the average contracted and demanded power, and the enhancement indicators, which will be defined in Section II.E. The structure of the customer dataset and the file containing the NACE-location pairs is depicted in Fig. 2. A more detailed description of how to obtain these files can be found in [33].

It should be noted that the demand profiles defined in this research, whether for individual customers or for NACE-location pairs, are not suitable for use as demand response (DR) baselines. Firstly, due to their low time-resolution, since they are characterized with one value for each month, whereas most DR applications require baselines with at least hour-by-hour values. Secondly, DR baselines should certainly model an individual customer, whereas the profiles of NACE-Location pairs are just median values of an aggregated demand.

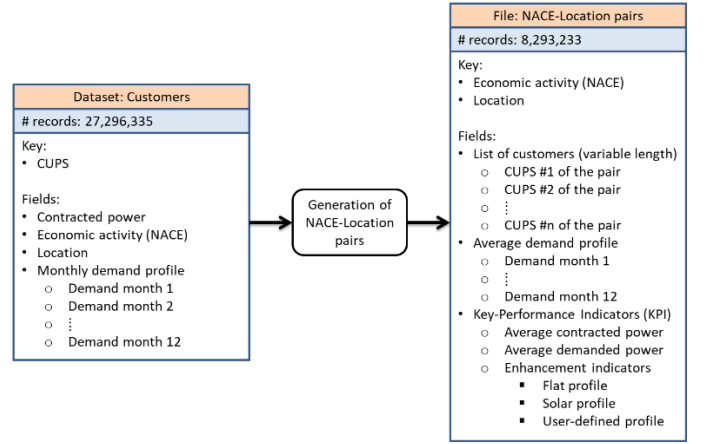

Fig. 2. Structure of the customer dataset and the file containing the NACE-location pairs.

### B. Target profiles

In this research, a particular target (or desired) demand profile, usually corresponding to the power generation (or acquisition) profile of the retailer, can be defined for every NACE-location pair. The target corresponding to the $k$-th pair is formalized using a vector $\boldsymbol{g}^{(k)} = \left\{ g_j^{(k)} \right\}$, where $j$ indicates the month of the year.

The general case regards a possible different target profile for every pair, allowing specific considerations for certain locations or types of customer. However, in most cases it is required a single target profile, corresponding to that with the best fit to the power generation (or acquisition) profile.

Up to three types of target profiles were explored through this paper. The first one is called the flat profile where the demand is the same in every month and for every pair. Therefore, a customer profile that is similar to this target is indicative that it has little variation throughout the year. It can be formally defined by the expression

$$\boldsymbol{g}^{(k)} = \left\{ g_j^{(k)} \right\}: \forall j, k, g_j^{(k)} = 1. \quad (1)$$

The second target, called solar profile, tries to accommodate the electricity demand to the solar radiation at the customer's location. Strictly speaking, there would be as many solar profiles as locations (8131 in our case). However, to simplify the problem, it is assumed that all the locations in the same province have the same solar radiation profile, thus obtaining only 52 of these profiles. Some examples of provincial solar profiles, as obtained from [34], are shown in Fig. 3.

Finally, an arbitrarily user-defined target profile is also considered. As an example, the aggregated power demand of a certain medium-size Spanish retailer (Medina-Garvey) is considered. Specifically, the overall demand during three years is shown in Fig. 4. Also its mean value (profile) is depicted, which is formally denoted as a vector $\boldsymbol{m} = \{m_j\}$, where $m_j$ is the value of the mean demand corresponding to the $j$-th month. Further, the mean aggregated demand over the year is defined as

$$\bar{m} = \frac{1}{12} \sum_{j=1}^{12} m_j. \quad (2)$$

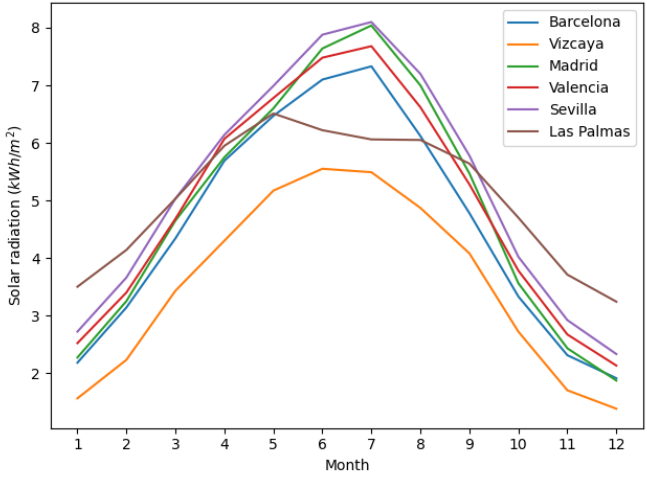

Fig. 3. Examples of solar radiation profiles at six different Spanish provinces.

This target profile corresponds to the consumption required to complement the overall consumption of the retailer, in such a way that the customers with this profile help to flatten the retailer's load curve. To obtain this result, the aggregated demand $\boldsymbol{m}$ is inverted with respect to its mean $\bar{m}$, where the inverted value corresponding to the $j$-th month is $w_j = 2\bar{m} - m_j$. This complementary profile $\boldsymbol{w}$ is also depicted in Fig. 4. It can easily be derived that its mean value is $\bar{w} = \bar{m}$. Eventually, the user defined target profile is obtained as the normalized version of $\boldsymbol{w}$:

$$\boldsymbol{g}^{(k)} = \left\{g_j^{(k)}\right\}: \forall k, g_j^{(k)} = \frac{w_j}{\bar{m}} = 2 - \frac{m_j}{\bar{m}}. \quad (3)$$

As can be seen, the aggregated demand $\boldsymbol{m}$ and its reverse $\boldsymbol{w}$ certainly yields a flat consumption.

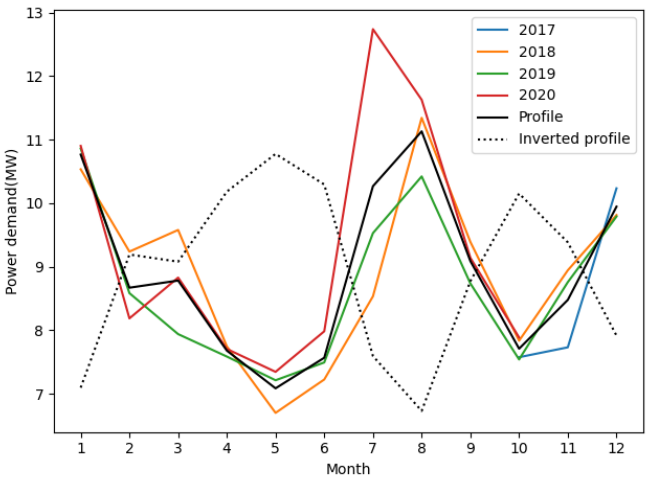

Fig. 4. Aggregate electricity demand profile of the customers of Medina-Garvey (a medium-size Spanish retailer).

### C. Distance between profiles

To establish a marketing strategy, it is previously convenient to measure how far a certain demand profile is from the target profile. For that reason, the concept of distance between two profiles is formalized in this Section.

The consumption profile of the $i$-th customer belonging to the $k$-th NACE-location pair is defined by the vector $\boldsymbol{c}^{(k,i)} = \left[c_1^{(k,i)}, c_2^{(k,i)}, \cdots, c_{12}^{(k,i)}\right]$, where $i$ is in the range $[1, n_k]$, and $n_k$ is the number of clients associated to that pair. This vector can also be defined as $\boldsymbol{c}^{(k,i)} = \left\{c_j^{(k,i)}\right\}$, where $j$ indicates the month of the year, which is in the range $[1..12]$.

The difference between this demand profile and the target profile is defined by the vector

$$\boldsymbol{\delta}^{(k,i)} = \left\{\delta_j^{(k,i)}\right\} = \boldsymbol{c}^{(k,i)} - \boldsymbol{g}^{(k)}. \quad (4)$$

This vector can be summarized by a single value as the Root Mean Square Distance (RMSD) of the differences corresponding to each month, that is, by

$$d^{(k,i)} = \text{RMSD}[\boldsymbol{\delta}^{(k,i)}] = \sqrt{\frac{1}{12} \sum_{j=1}^{12} \left(c_j^{(k,i)} - g_j^{(k)}\right)^2}. \quad (5)$$

To avoid outliers in energy consumptions, the average distance for each $k$-th NACE-location pair is defined using the median, instead of the mean, that is by

$$d^{(k)} = \underset{i \in [1..n_k]}{\text{median}} d^{(k,i)}. \quad (6)$$

Besides, the electricity consumption of the $i$-th customer belonging to the full dataset can be defined defined by the vector $\boldsymbol{c}^{(*,i)}$. This customer has a certain economic activity (NACE) and is situated in a particular location, so it also belongs to a specific NACE-location pair, identified as the $k_i$-th pair. Then, the difference between this consumption and its target can be written as $\boldsymbol{\delta}^{(*,i)} = \left\{\delta_j^{(*,i)}\right\} = \boldsymbol{c}^{(*,i)} - \boldsymbol{g}^{(k_i)}$, and the corresponding distance is

$$d^{(*,i)} = \text{RMSD}[\boldsymbol{\delta}^{(*,i)}] = \sqrt{\frac{1}{12} \sum_{j=1}^{12} \left(c_j^{(*,i)} - g_j^{(k_i)}\right)^2}. \quad (7)$$

Thus, the average distance for the entire dataset is defined by

$$d^{(*)} = \underset{i \in [1..n]}{\text{median}} d^{(*,i)}, \quad (8)$$

where $n$ is the total number of customers.

### D. Enhancement associated to a customer

Let us define the marketing strategy $\mathcal{S}^{(k,i)}$, as the one that identify the $i$-th customer belonging to the $k$-th NACE-location pair as the next customer to be attracted. Then, the demand profile of that customer is at a distance $d^{(k,i)}$ from the target profile $\boldsymbol{g}^{(k)}$, as defined in (5).

Besides, let us call the lack of any marketing strategy, or null strategy $\mathcal{S}^{(*)}$, as the one where the next attracted customer is randomly selected from the entire dataset. Then, its average distance from the target profile will be $d^{(*)}$, defined by (8)

So, to gauge the advantage of using the marketing strategy $\mathcal{S}^{(k,i)}$ versus the null strategy $\mathcal{S}^{(*)}$, an enhancement metric can be defined according to the expression

$$e^{(k,i)} \equiv \frac{d^{(*)} - d^{(k,i)}}{d^{(*)}} = 1 - \frac{d^{(k,i)}}{d^{(*)}}. \quad (9)$$

This gauge has a positive value if the $\mathcal{S}^{(k,i)}$ marketing policy obtains a new customer who has a distance to its target which is less than the difference that would be obtained by not applying any marketing policy, $\mathcal{S}^{(*)}$. Otherwise, the metric will be negative.

For most of the customers, the values of this enhancement metric are in the range $[-1, +1]$. As the distances are, by definition, positive values, the upper bound of the metric is $e^{(k,i)} \leq +1$. However, the metric is not lower-bounded and a few values below $-1$ can be found.

To overcome this situation the enhancement metric $e^{(k,i)}$ is transformed into an enhancement indicator $E^{(k,i)}$, which is obtained using the expression

$$E^{(k,i)} = \text{EID}[e^{(k,i)}] = \begin{cases} 1, \forall e^{(k,i)} > 1 \\ e^{(k,i)}, \forall 0 \leq e^{(k,i)} \leq 1 \\ \exp[e^{(k,i)}] - 1, \forall e^{(k,i)} < 0 \end{cases}. \quad (10)$$

This normalizing transformation is also depicted in Fig. 5. It can be seen that $E^{(k,i)} = e^{(k,i)}$ for metrics in the range $[0, +1]$, while the indicator's absolute value is never greater than 1.

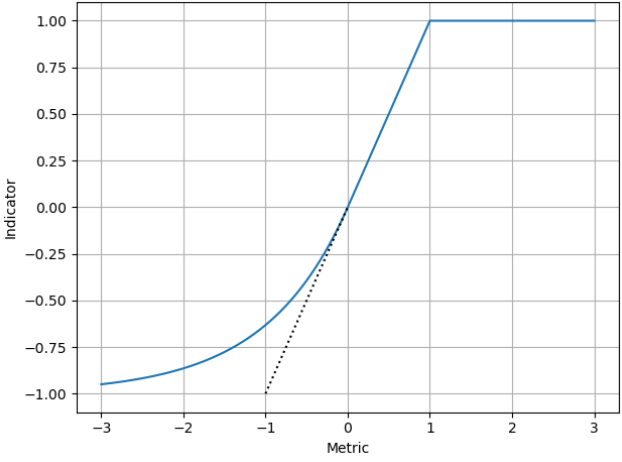

Fig. 5. Transformation of the enhancement metric into the enhancement indicator.

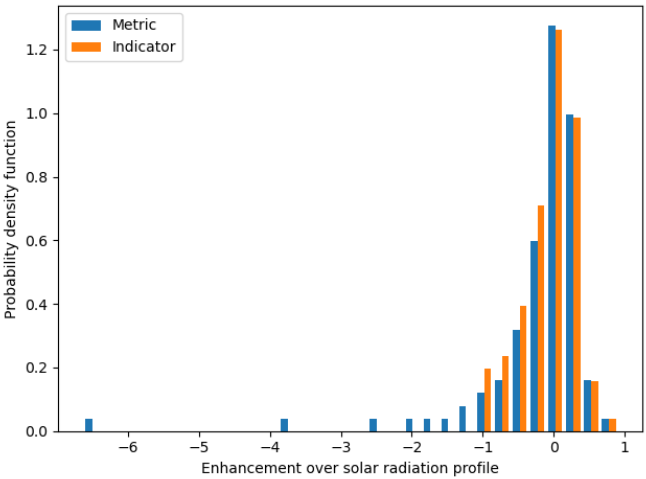

Fig. 6. Histograms of the enhancement metrics (blue) and indicators (orange) corresponding to the entire dataset.

Let us consider, for example, the values of the enhancement metric for the entire dataset when a solar radiation profile is used as the target profile. The histogram of these values is depicted (in blue) in Fig. 6.

As can be seen, about a 90% of the profiles have enhancement metrics in the range $[-1, +1]$. Approximately a 5% of profiles have values in the range $[-2, -1]$, and the remaining 5% of the profiles have values below $-2$. After the normalization transformation (10), the enhancement indicators are distributed according to the histogram shown (in orange) in Fig. 6. As can be seen, only the extreme original values are significantly modified.

The customer-based marketing strategy assumes that the most convenient customer to be attracted is that with the smallest distance from the target profile, that is, with the highest enhancement indicator.

### E. Enhancement associated to a NACE-location pair

To define a marketing strategy associated to a specific customer requires that certain identifying information be available, such as the name, postal or e-mail address, phone number, etc. Then, once determined the demand profile with the highest enhancement potential, their associated identifying data is used to contact the prospective customer and to offer him/her a personalized and more attractive tariff.

However, the open datasets available to foster competition among retailers (as was defined in Section II:A) only contain anonymized demand information. Then, marketing strategies associated to attract a particular customer have to be dismissed. Alternatively, as these datasets link the demand profile of each customer to its economic activity and location, the marketing strategy can be established using this information.

For that purpose, let us define the marketing strategy $\mathcal{S}^{(k)}$, as the one that randomly attracts the next customer among those belonging to the $k$-th NACE-location pair.

Then, the concept of enhancement metric of a customer must be extended to NACE-location pairs. Indeed, by applying the marketing strategy $\mathcal{S}^{(k)}$, the next customer will have a demand profile that, on the average, will be at a distance from the target profile, $d^{(k)}$, described in (6). Although that is the standard definition of distance, its application to the profile of a pair (rather than to the profile of an individual customer) has not previously proposed in the literature.

Therefore, the enhancement metric of the marketing policy $\mathcal{S}^{(k)}$ that prioritizes the $k$-th pair can be defined according to the expression

$$e^{(k)} \equiv \frac{d^{(*)} - d^{(k)}}{d^{(*)}} = 1 - \frac{d^{(k)}}{d^{(*)}}. \quad (11)$$

Thus, the corresponding enhancement indicator is then obtained by the expression

$$E^{(k)} = \text{EID}[e^{(k)}]. \quad (12)$$

The proposed marketing strategy assumes that the most convenient customer to be attracted is randomly selected from those belonging to the NACE-location pair with the smallest distance from the target profile, that is, with the highest enhancement indicator.

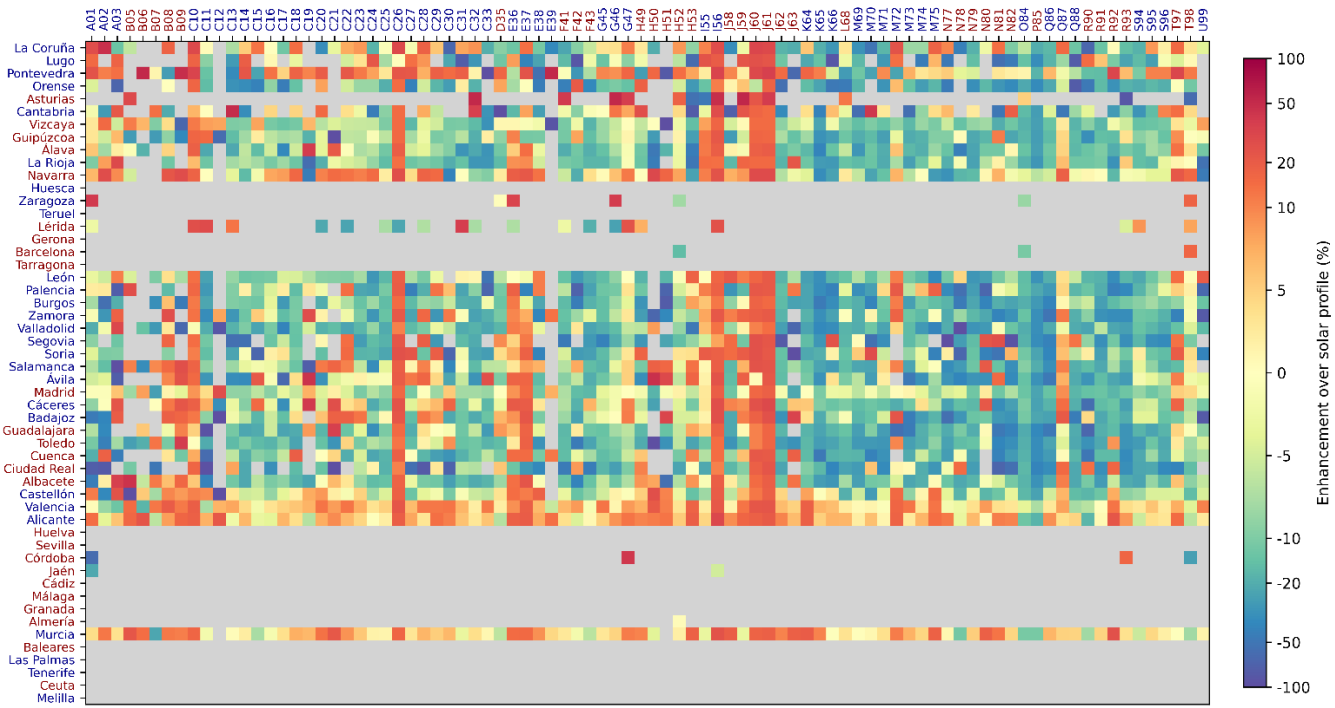

Fig. 7. Matrix representation of the enhancement indicators over solar profile (in %).

## III. RESULTS

Enhancement indicators have been computed for every NACE-Location pair and can be presented in different ways. As example, an overview of their values may adopt the form of a matrix, where there is a row for each location, a column for each economic activity (NACE), and the color of each cell denotes the corresponding enhancement indicator. As the size of the matrix is too high (8203 x 1011), it is usually represented aggregating locations by provinces (up to 52 values) and NACE codes by division (88 values).

The resulting matrix (in %), using for example a solar profile as the target, is depicted in Fig. 7. Following this general overview, an electricity retailer should be advised to concentrate its marketing efforts in, for instance, the 16,692 customers with NACE code J61 (telecommunications), or those located at the province of Alicante (1,035,476 customers).

The cells colored in gray correspond to NACE-location pairs where there is no customer. Usually, this situation occurs in certain provinces where the records about customers do not contain the NACE information.

The enhancement indicators may also be used to advise an electricity retailer of the best locations to invest marketing efforts. This is equivalent to compare the rows of the matrix in Fig. 7.

For example, the distribution of values for the enhancement indicators (in %) at two provinces (Albacete and Pontevedra), using a solar profile as the target, are shown in Fig. 8. As can be seen in this figure, an average customer in Pontevedra will have an electricity demand closer to the solar profile than the average customer in Albacete.

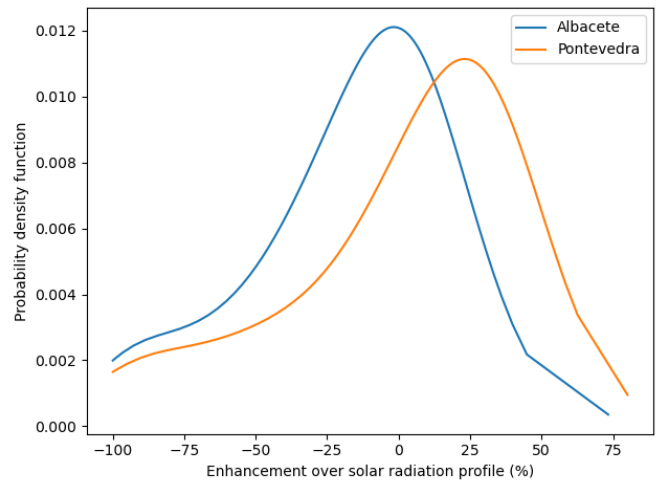

Fig. 8. Probability density function of the enhancement indicators over solar profile (in %) at Albacete and Pontevedra provinces.

Finally, the enhancement indicators may be used to advise an electricity retailer, operating at a certain location, of the best economic sectors to invest marketing efforts.

This is equivalent to explore by columns a single row of the matrix in Fig. 7. For example, the values for the enhancement indicators (in %) at the province of Pontevedra (location code: P36), using a solar profile as the target, are shown in Fig. 9 as a pie chart. The sectors colored in gray correspond to NACE codes with no customer at this location.

It is important to highlight that, although the used datasets do not identify a particular customer, some NACE-location pairs contain so few customers (even one single customer) that is not difficult their indirect identification. Indeed, about a 96% of the pairs are empty, that is, only a 4% of the pairs have one or more customers. And, among these non-empty pairs, a 50% have one customer, while about a 90% have 10 or fewer customers (see Fig. 10).

Converting these results in terms of the number of customers who can be indirectly identified, it can be said that 93,176 customer are uniquely identified, while 284,804 are identifiable in sets of 10 or less customers.

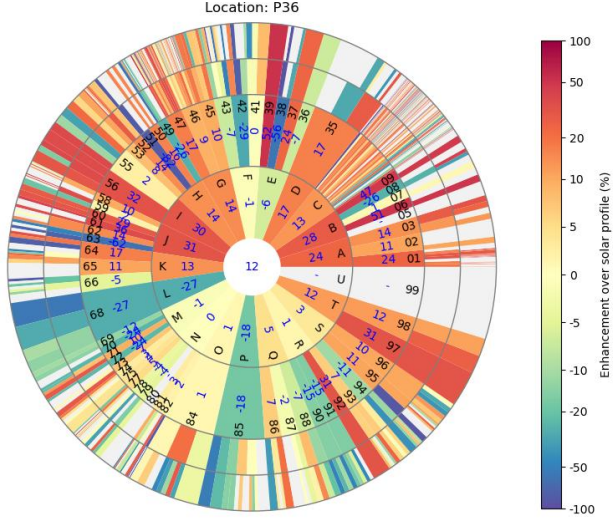

Fig. 9. Distribution by economic activity (NACE codes) of the enhancement indicators over solar profile (in %) at the province of Pontevedra.

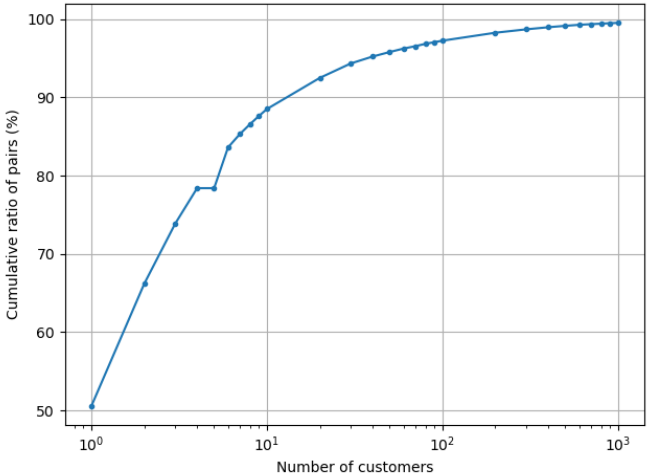

Fig. 10. Ratio of non-empty NACE-location pairs containing up to a certain number of customers.

## IV. DISCUSSION

To assess the marketing strategy proposed in the methodology section, a simulation process has been carried out. It has been measured the distance of the aggregate demand from a certain target profile, as seen by a simulated retailer which is sequentially attracting new customers using the described marketing strategy. The process begins selecting the NACE-location pair with the highest enhancement indicator and incorporating their customers in a random order. Once all the customers of that pair have been attracted, the simulation selects the pair with the second highest enhancement indicator, and then, their customers are randomly incorporated. The process continues selecting pairs ordered by decreasing enhancement indicators, and randomly incorporating their customers.

The proposed methodology is evaluated by obtaining the aggregate demand at each step, whenever a new customer is incorporated. The profile of this accumulated demand is compared to the target demand and the distance between these two profiles is obtained and plotted for an increasing number of new customers. The result, using for example a solar profile as the target, is depicted (blue line) in Fig. 11.

To obtain that evolution for 1 million of new customers, a processing time of 1570 seconds (about 26 minutes) has been required, running on a desktop computer based on an Intel® Core™ i7-11700 @ 2.50GHz processor, with 64 GB of RAM and a Solid State Disk (SSD) of 2TB, and using a script developed in Python 3.9 under a Windows 11 operating system. This processing time is linearly related to the number of customers.

To evaluate the proposed strategy, it is first compared to the case where the customers are being incorporated into the aggregate demand in a random order, that is, without following any marketing strategy.

In this case a random sample of 1,000,000 clients is generated from the dataset of customers (27,296,335 records). Then the clients in the sample are sequentially incorporated obtaining a corresponding aggregate demand at each step. As for the proposed marketing strategy (previously described), the profile of the randomly accumulated demand is compared to the target profile and their difference is obtained as a function of the number of customers aggregated. The described procedure takes about one minute (62 seconds) using the previously described computer. To avoid a special (extreme) result from a particular realization of the randomness, this random process is repeated 100 times and the median value in each step is considered.

As can be observed, the aggregate demand using the proposed marketing strategy ($d_m$, blue) are much closer to the target profile than that obtained by the median value of the random incorporation of new customers ($d_r$, in red in the figure). The interquartile range (IQR) of the distance to the target profile using this random selection of clients is also depicted as a light red shadow.

Besides the enhance indicators, every NACE-location pair contains information of the average contracted and demanded power of all the customer belonging to that pair. For the sake of validation, these values are also used as two new competing marketing strategies. In these cases, the customers with a higher contracted power (or a higher demanded power) are prioritized. The resulting evolutions are also depicted in Fig. 11.

Therefore, the proposed strategy clearly improves the adjustment of aggregate demand to the target profile. For example, adding one thousand new customers yields an aggregate demand which is at distance of $d_m = 0.209$ from the solar profile, while the distance would be $d_r = 0.474$ if the customers were randomly incorporated. This means that the proposed marketing strategy reduces the distance from the

target by a 56%. This relative reduction is obtained using the expression

$$R = \frac{d_r - d_m}{d_r}. \quad (13)$$

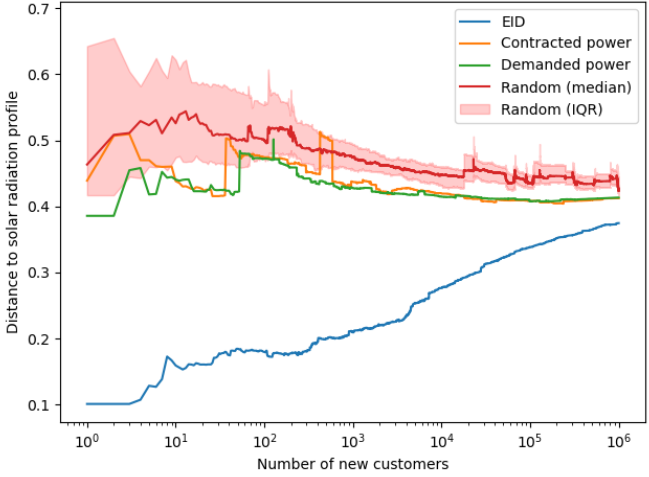

Fig. 11. Distance of the aggregate demand from the solar radiation profile for an increasing number of new customers using several strategies to select them: enhancement indicator (EID), contracted power, demanded power and random selection, showing the median and the interquartile range (IQR).

If, alternatively, the contracted (or demanded) power is used as the customer selection criterion (adding one thousand new clients), the distance to the target is reduced by only 10% compared to random selection, confirming the advantage of the proposed strategy.

The reduction provided by the proposed marketing strategy decreases as new customers are added, as the best ones are considered first. The relationship of this reduction to the number of new customers is depicted in Fig. 12 for the three target profiles described in Section II.B. It can be seen that, even for a large number of new customers, the proposed marketing strategy obtains a significant reduction of the distance, that is, it yields to an aggregate demand which is significantly closer from the target profile.

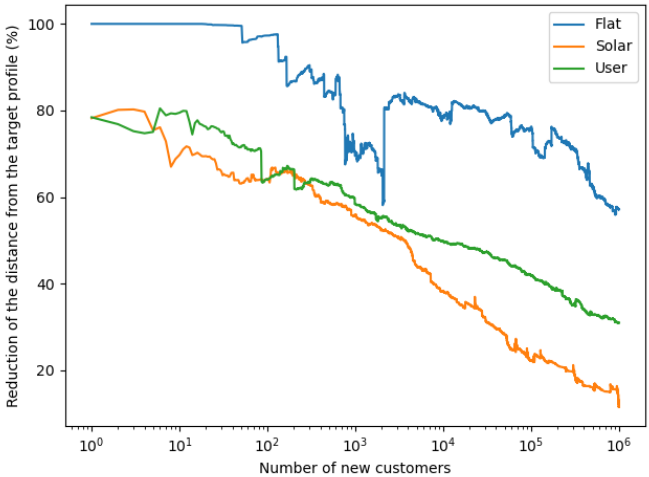

Fig. 12. Reduction of the distance of the aggregate demand from the target profile for an increasing number of new customers.

The higher reductions in the case of using as target a flat profile are because some of the new customers have a demand which is flat or almost flat ($d_m \approx 0$), that is, much closer from the flat profile than to the solar or user-defined profile. Then, according to (13), the corresponding relative reduction is $R \approx 1$.

It must be remarked that the electricity retailers in Spain have, on the average, about 100,000 ($10^5$) customers (although the market is not equally distributed, and many retailers have much less customers). According to the previous results, adding a 10% of new customers ($10^4$) can be done with a reduction of distances from the target profile of about 40%. Even in the case of doubling the customer base ($10^5$ new clients), reductions of about 30% can still be expected.

## V. CONCLUSION

This paper has explored the use of semi-public electricity demand databases to improve marketing strategies for energy retailer companies. As these databases do not contain enough information to identify a particular customer, the marketing decisions are based on data about the economic sector (NACE) and location associated to every demand profile.

The use of the enhancement indicator (a KPI based on the distance of the demand profile from a certain target profile), have shown a valuable gauge to lead marketing efforts.

The proposed marketing strategy has significantly reduced the distance from the target profile, in comparison to randomly adding new customers. Adding 10,000 new customers can be done with a 40% reduction, while adding 100,000 still are possible with a 30% decrease.

The pair NACE-location data, which are available in the datasets, has been revealed as a powerful tool for indirect identification of customers. Indeed, about 100,000 customers are uniquely identified by their corresponding pairs, while about 300,000 customers are identifiable in sets containing 10 or less consumers.

Thus, it is important to highlight as novelty of the proposed work, the large number of customers analyzed (more than 27 millions), the grouping of these customers by pairs attending to their economic activity and location, the large number of pairs employed (about half a million), the introduction of the enhancement indicator associated to particular customers and to NACE-location pairs, and finally, the definition of a marketing strategy for retailing companies.

## VI. ACKNOWLEDGMENT

The authors thank the TSO Company and Medina-Garvey for granting access to their datasets.

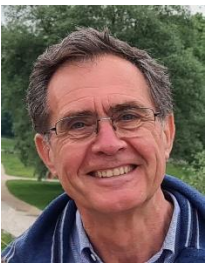

**Joaquin Luque** (SM'20) received his M.S. (1980) and Ph.D. (1986) degrees in industrial engineering (electrical engineering) and an M.S. degree in philosophy (1994) at the University of Seville, Spain. He is currently a Full Professor of electronic engineering at the University of Seville, with more than 30 years of teaching and research experience in different computer engineering disciplines mainly related to basic electronics, communications and control. He also has extended managing experience at the University of Seville, where he has been Head of Department (1993-2000) and Rector (2008-2012). Dr. Luque has been a Visiting Scholar at the University of California Berkeley, USA (2018), and an Invited Professor at the University of Genoa, Italy (2019).

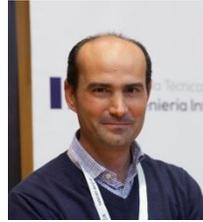

**Alejandro Carrasco** (M'07) received his Computer Engineering degree in 1998 and his Ph.D. in Computer Engineering in 2003 from the University of Seville (Spain). Since 1997, he has worked for several companies in the area of software engineering and computer networks, has founded a new technology-based company (NTBF) and he has actively participated and directed several R&D projects. Currently Dr. Carrasco is a Lecturer and Researcher at the Department of Electronic Technology at the University of Seville, and his research activities are focused on data mining, distributed services and artificial intelligence applied to industrial computing and cybersecurity.

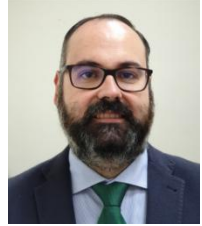

**Enrique Personal** (S'10-M'16) ) received his degree in industrial electronic engineering and his degree in automatic control and industrial electronic engineering from the University of Seville, Spain, in 2006 and 2009, respectively, where he also obtained his Ph.D. degree in industrial computer science in 2016. He is currently an Associate Professor in the Department of Electronic Technology at the University of Seville. His main fields of interest are smart grids, fault location methods, power systems, demand-side management, and flexibility.

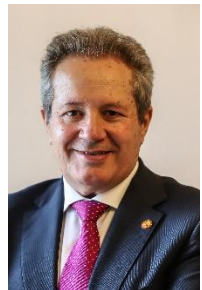

**Francisco Pérez** received his B.Sc. degree in Electronic Physics and the Ph.D. degree in Robotics and Digital Imaging Processing from the University of Seville, Spain, in 1985 and 1992, respectively. He has been teaching Digital Electronics, Computer Architecture, and Computer Networks for more than 30 years, in the School of Computer Science and Engineering at the University of Seville, where he was Dean (1996-2006). He is, currently, a Full Professor and the Head of the Department of Electronics Technology. His research areas include industrial communications, embedded control devices, IoT, smart grids, and smart cities.

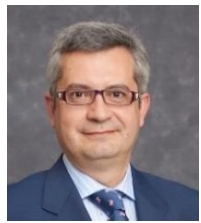

**Carlos Leon** (SM'10) received the B.Sc. degree in electronic physics and the Ph.D. degree in computer science from the University of Seville, Seville, Spain, in 1991 and 1995, respectively. He is currently a Full Professor of electronic engineering and computer science at the University of Seville and Head of the Telefonica Chair. His research areas include knowledge-based systems, computational intelligence, big data analytics, blockchain, edge computing, cyber physical IoT systems and machine learning, focus on Utilities System Management. On these topics he is authors of more 200 papers and conference contributions. He has been director or principal investigator of more than 70 research projects, mainly in collaborations with companies. He is a Senior Member of the IEEE Power Engineering Society.